\documentclass[aps,twocolumn,showpacs,prb]{revtex4-1}
\usepackage[english]{babel}
\usepackage{amsmath}
\usepackage{graphicx} 
\usepackage{times}
\usepackage{epsfig} 

\usepackage{color}
\definecolor{nred}   {RGB}{224,0,0}
\definecolor{nblue}  {RGB}{28,130,185}
\definecolor{dgreen} {RGB}{78,138,21}
\definecolor{norange}{RGB}{230,120,20}

\begin{document} 

\title{Decay of density waves in coupled one-dimensional many-body-localized systems}

\author{Peter Prelov\v{s}ek$^{1,2}$}
\affiliation{$^1$Jo\v zef Stefan Institute, SI-1000 Ljubljana, Slovenia}
\affiliation{$^2$ Faculty of Mathematics and Physics, University of
Ljubljana, SI-1000 Ljubljana, Slovenia}


\begin{abstract}
The behavior of coupled disordered one-dimensional systems, as modelled by identical fermionic 
Hubbard chains with the on-site potential disorder and coupling emerging through the inter-chain hopping $t'$, is analysed. 
The study is motivated by the experiment on fermionic  cold atoms on a disordered lattice, where a decay rate
of the quenched density wave was measured. We present a derivation of the decay rate $\Gamma$
within  perturbation theory  and show that even at large disorder along the chains 
the interaction leads to finite $\Gamma > 0$, the mechanism being the interaction-induced coupling
of in-chain localized and inter-chain extended single-fermion states. Explicit expressions for $\Gamma$
are presented for a weak interaction $U < t,t'$, but extended also to the  regime $t > U >t'$. It is shown that in 
both regimes  $\Gamma$ increases with the inter-chain hopping $t'$, as well as decreases with increasing disorder. 
\end{abstract}


\maketitle

\section{Introduction}
The paradigm of the many-body localization (MBL)
represents the extension of well understood single-particle Anderson localization 
\cite{anderson58,mott68,kramer93}  to fermionic systems with a repulsive interaction.  While original proposals
for the MBL state were dealing with systems  with a weak disorder \cite{fleishman80,basko06}, by now numerous 
theoretical studies confirm the existence of a  MBL-like state in
the regime of strong disorder and moderate interactions. Most studies so far were performed by the numerical investigation
of the prototype model, being the one-dimensional (1D) model of disordered interacting spinless fermions, equivalent to the anisotropic 
Heisenberg chain with random local fields,  \cite{oganesyan07,znidaric08,monthus10,berkelbach10,barisic10,pal10,bardarson12,
serbyn13,bera15,luitz15,agarwal15,lazarides15}. Results confirm that for large disorder
$W > W_c$ systems reveal some basic features of the MBL, referring here to those relevant
also for experiments : a) the absence of d.c. transport at any temperature $T$ 
\cite{berkelbach10,barisic10,agarwal15,gopal15,lev15,steinigeweg15,barisic16,kozarzewski16},
b) generally nonergodic behavior of correlation functions and of quenched initial quantum states 
\cite{pal10,serbyn13,lev14,schreiber15,serbyn15,khemani15,luitz16}, c) the area law instead of
volume law for entropy, still with a logarithmic growth in the MBL phase \cite{znidaric08,bardarson12,kjall14,serbyn15,luitz16}. 
Even within apparently simple 1D models there are essential theoretical and numerical challenges, 
among them also the nature of the MBL transition, e.g., well defined 
phase transition \cite{pal10,serbyn15,vosk14,vosk15,potter15} vs. sharp crossover \cite{barisic10,barisic16}, 
and the prediction of measurable signatures of the MBL transition. 

On the other hand, cold atoms in optical lattices have already provided a direct experimental insight 
into the MBL phenomenon and have shown the qualitative transition
between an ergodic and the nonergodic  phase. Studies of 1D disordered systems of cold atoms 
\cite{schreiber15,kondov15} have been recently extended to
coupled 1D systems \cite{bordia16} as well to systems with a full two-dimensional (2D) disorder \cite{choi16}. The 
motivation for this work is the former experiment which clearly reveals that in coupled 
chains of localized fermions with identical disorder, the fermion interaction $U$ leads to the
decay and the thermalization of the initial density-wave (DW) state. This implies also that the 1D nonergodic
behavior is destroyed in the presence of $U \neq 0$ by the inter-chain coupling provided that
there is no inter-chain disorder. Such an observation and its understanding may be very important for 
further explorations of the MBL physics in higher dimensions. Theoretically, there are few studies 
discussing MBL physics beyond 1D, e.g. in ladders \cite{baygan15} and in 2D systems \cite{barlev16}. 
We should also note that for cold-atom systems the appropriate model
is the Hubbard model, which is much less explored with respect to possibility of MBL physics \cite{mondaini15,
reichl16,barlev16}
and might even reveal some qualitative differences (taking into account additional symmetries \cite{potter16}) relative 
to prototype disordered spinless models predominantly studied so far.  

In this paper we show on the example of coupled identical disordered Hubbard chains
that the decay mechanism of the initial out-of-equilibrium state is related to the Hubbard interaction $U$,
coupling the in-chain localized and inter-chain extended single-particle states. In particular, we
formulate the analytical procedure for the calculation of the decay of an initial density-wave (DW) state, 
as relevant  for cold-atom experiment \cite{bordia16}. In the latter the measured quantity is time-dependent imbalance 
$I(\tau)$. The breaking of the ergodicity of the latter, i.e., $I(\tau\to \infty) >0 $
 can be considered as a measurable order parameter for the nonergodic state.  We do not address here in more
detail the possible (or at least slow) decay of initial DW state in uncoupled chains. We show, however, that the 
inter-chain coupling introduces even for large disorder  a relevant and leading additional decay channel for DW 
decay.

In Sec.~II we present the model and its representation within the basis of 1D localized states. We introduce
also the relevant DW operators studied further on. Sec.~III is devoted to the derivation of the DW decay rate
within the perturbation theory, leading to an approximation in terms of a Fermi-golden-rule expression.
Sec.~IV presents results within the perturbative regime $U \ll t,t'$ for the DW rate for the case of coupled
chains, touching also the relation to the problem of 1D DW decay and possible  generalizations.
Conclusions are given in Sec.~V.  
  
\section{Model}

To remain close to the experiment \cite{bordia16} we consider in the following the (repulsive) 
fermion Hubbard  model on coupled chains where the disorder is identical in all chains, 
\begin{eqnarray}
H &=& \sum_j H_{0j} - t' \sum_{ljs }( c^\dagger_{l,j+1,s} c_{lj,s} + \mathrm{H.c.})  + H_U , \nonumber \\
H_{0j} &=& - t \sum_{ls } ( c^\dagger_{l+1,j,s} c_{lj,s} + \mathrm{H.c.} ) +  \sum_l h_l n_{lj}  \nonumber \\
H_U&=& U \sum_{lj} n_{lj\uparrow} n_{lj\downarrow}, \label{hub}
\end{eqnarray}
with the in-chain (site index $l$) and inter-chain (chain index $j)$ nearest-neighbor (n.n.) hopping $t,t'>0$, respectively. 
$n_{lj}=\sum_s n_{ljs}$  and we assume the disorder entering via  random and 
independent local potentials $-W < h_l <W$, the same in all chains. We note that within the actual experiment 
\cite{bordia16} $h_l$ are quasi-random.  For further analysis it is relevant
that we consider the filling $\bar n < 1$ (in the actual experiment $\bar n \sim 1/2$), 
avoiding the scenario of an (Mott) insulating state entirely due to repulsive $U >0$. Further on we also 
consider only the case of weaker inter-chain hopping $t' < t$.

 Let us start by considering a single 1D chain as described by $H_{0j}$ in Eq.~(\ref{hub}),
where we omit for simplicity the index $j$. One can find first single-particle eigenfunctions of $H_{0}$ which are 
localized states for $W>0$,  
\begin{equation}
 |\phi_{ms} \rangle = \varphi^\dagger_{ms} |0\rangle =  \sum_l \phi_{ml} c^\dagger_{l s} |0\rangle,
 \quad H_0 = \sum_{ms} \epsilon_m \tilde n_{ms},
\end{equation}
where $\tilde n_{ms}$ is the occupation of the 
single-particle localized  state. One can then represent $H_U$ in terms of such localized states,
\begin{eqnarray}
H_U&=&  U \sum_{mm'nn'} \chi_{mn}^{m'n'} ~\varphi^\dagger_{m'\uparrow} \varphi^\dagger_{n'\downarrow} 
\varphi_{n\downarrow} \varphi_{m\uparrow} , \nonumber \\
\chi_{mn}^{m'n'} &=& \sum_{l} \phi_{m'l} \phi_{n'l} \phi_{nl} \phi_{ml}, \label{vklmn}
\end{eqnarray}
where coefficients $\chi_{mn}^{m'n'}$ are by construction invariant on the  index permutation, and indices 
$m,m',n,n'$ further on refer to 1D localized basis, ordered conveniently by the position of the maxima of localized functions.

Let us consider many-body (MB) states  
$| {\underline m}\rangle = \prod_{\underline m} \varphi_{ms}^\dagger |0\rangle $
within such a localized basis. In this representation one term is  the diagonal (Hartree-Fock) correction 
\begin{equation}
H'_{d}= U \sum_{mn} \chi_{mn}^{mn}  \tilde n_{n\uparrow} \tilde n_{m\downarrow}, \label{hd}
\end{equation}
so that we can separate $H_U=H'_d+H^{\prime\prime}$, and only $H^{\prime\prime} \neq 0$
can mix different $| {\underline m}\rangle$.

Our goal is the behavior of the staggered DW operator, defined by
\begin{equation}
A= \sum_{l} (-1)^l n_{l} /\sqrt{L}, \label{dwa}
\end{equation}
In particular, we wish to follow its time dependence, being directly related 
to the measured imbalance $I(\tau) \propto \langle A \rangle (\tau)$ emerging from an initial state  
$\langle A \rangle (\tau = 0) \neq 0$.
Starting in experiment \cite{schreiber15,bordia16} as well as in numerical studies  \cite{schreiber15,mondaini15}, 
with a DW eigenstate $A |\Psi_0 \rangle = A_0 |\Psi_0 \rangle$, leads to fast initial dynamics (including oscillations) on the 
timescale $\tau \sim 1/t$, representing the decomposition of $|\Psi_0 \rangle$ into diferent localized $| {\underline m}\rangle$.
We are rather interested in long-time decay, beyond the former short-time transient, which is qualitatively of the 
form $I(\tau)= I_0(\tau) \mathrm{exp}(-\Gamma \tau)$.
In particular, we study decay-rate $\Gamma$ emerging from the dominant channel due to the interchain 
coupling, as appears also in the experiment \cite{bordia16}. For such long-time decay it is more convenient to analyze 
the modified DW operator, given already in terms of localized states,
\begin{equation}
B = \frac{1}{\sqrt{L}}  \sum_{ms} (-1)^m \tilde n_{ms}. \label{dwb}
\end{equation}
We can for convenience assume that localized states are ordered by the site $m$ where they have 
maximum amplitude.  It is evident that in the case $H^{\prime\prime}=0$, the initial state 
$| {\underline m}\rangle$ would not decay as well as $\langle B (\tau)\rangle $ would be constant, in contrast 
to more standard definition via Eq.~(\ref{dwa}). 

\section{Density-wave decay rate: derivation}

The goal is to evaluate  $\langle B(\tau)\rangle $ when perturbed from the initial value $\langle B\rangle_0=0$.
In actual experiment the deviation can be and actually is large \cite{bordia16}. Still we assume that the system 
under consideration (as well as in experiment \cite{bordia16}) is  ergodic and approaches the thermal 
equilibrium. Final DW decay rate should be therefore determined by the equilibrium and  consistent with an analytical 
approach to the problem we therefore  apply the linear-response theory for the DW decay
to the equilibrium, as characterized by the temperature $T>0$ and the average particle density
$\bar n$.
The information is then contained within  susceptibility for the modified DW observable, i.e. 
\begin{equation}
\chi_{B}(\omega) =  -i\int_0^\infty \mathrm{e}^{i \omega  t} \langle[B(t),B]\rangle,
\end{equation}

To derive the expression for the DW decay rate $\Gamma$ within perturbation theory, 
as used e.g. for the  dynamical conductivity \cite{gotze72}, we follow 
the memory function formalism \cite{mori65,forster95,gotze72}, since it has the advantage 
to be easily extended  to nonergodic cases (as expeced within the MBL phase). 
Besides $\chi_{B}(\omega)$ we define in ususal way the relaxation function $\phi_B(\omega)$ \cite{mori65,forster95,gotze72} 
and static (thermodynamic) susceptibility $\chi_{B}^0$, 
\begin{equation}
\phi_B(\omega) = \frac{\chi_{B}(\omega)-\chi^0_{B}}{\omega}, \qquad
\chi_{B}^0 = \int_0^\beta  d\tau \langle B^\dagger B(i\tau) \rangle, \label{correl}
\end{equation}
where $\beta =1 /T$. In an ergodic case  $\chi^0_B=\chi_{B}(\omega \to 0)$, while in a nonergodic
system one has to consider also the possibility of $\chi^0_B> \chi_{B}(\omega \to 0)$.
Nevertheless, our study deals with the situation where (at least due to inter-chain coupling) there is 
a decay towards the equilibrium (thermalization).  
Due to general equlibrium properties of $\phi_B(\omega)$, we can represent it in terms
of the complex memory function \cite{gotze72}, 
\begin{equation}
\phi_B(\omega)= - \frac{\chi^0_{B}}{\omega + M(\omega)}.
\end{equation}
Skipping formal representation for the memory function $M(\omega)$ \cite{mori65,forster95}, we turn directly
to the  simplified expression valid within the perturbation theory \cite{gotze72}, 
\begin{equation}
M(\omega) = \frac{\chi_{F}(z)-\chi^0_{F}}{\omega \chi^0_{B}},
\end{equation}
where $\chi_{F}(z)$ is defined in analogy to Eq.~(\ref{correl}), for the operator $F =[H, B]=[H^{\prime\prime},B]$. 
The latter represents the effective force on the DW operator $B$,
\begin{equation}
F = \frac{2U}{\sqrt{L}} \sum_{mm'nn' s} \chi_{mn}^{m'n'} 
\zeta_{mm'}  \varphi^\dagger_{n',-s}   \varphi_{n,-s} \varphi^\dagger_{m' s} \varphi_{m s} , \label{f}
\end{equation}
where  $\zeta_{mm'}=0$ for even $m'-m$ and  $\zeta_{mm'}=(-1)^m $ for odd $m'-m$. 

Within the perturbation theory and within the 
eigenbasis of $H_0$  we further get
\begin{equation}
\chi_{F}(\omega) = - \frac{1}{Z} \sum_{\underline{n}, \underline {m}} \frac{ \mathrm{e}^{-\beta E_{\underline{n}} } -
\mathrm{e}^{-\beta E_{\underline{m}} } }{\omega+i(E_{\underline{n} }-E_{\underline{m}} )}
~|\langle \underline{n} |F| \underline{m} \rangle|^2 
\end{equation}
where $Z= \sum_{\underline{m}} \mathrm{e}^{-\beta E_{\underline{m}}}$. For the decay of interest is 
primarily the low-$\omega$ value $\Gamma = M(\omega \to 0)$ (provided that $M(\omega)$
dependence is modest) and for $\omega \ll T$ we obtain,
\begin{eqnarray}
\Gamma &=& \sum_{\underline{m}} p_{\underline{m}} \Gamma_{\underline{m} }, \nonumber \\
\Gamma_{\underline{m}} &=& \frac{\pi \beta }{ \chi^0_B }  \sum_{ \underline{n}} |\langle \underline{n}|F| \underline{m}\rangle|^2 
\delta (E_{\underline{n}} -E_{\underline{m}}),  \label{gamma}
\end{eqnarray}
where $p_{\underline{m}}= \mathrm{e}^{-\beta E_{\underline{m} }}/Z$ is the Boltzmann probability 
and $ \Gamma_{\underline{m} }$ are decay rates of particular states. We note that $\Gamma$ in Eq.~(\ref{gamma}
takes the simple form of generalized Fermi golden rule (FGR) for the considered problem. It should be noted that such a formulation, 
taking into account the form Eq.~(\ref{f}), also yields
$\Gamma$ as well as $\Gamma_{\underline {m}}$ as an intensive quantity, i.e. they do not depend of the system 
size $L$. 

Further simplification can be obtained for high $T$, i.e., where from Eq.~(\ref{correl}) we get
\begin{equation}
 \chi^0_{B} = \beta  \langle B^2 \rangle, \qquad \langle B^2 \rangle = \bar n(1-\bar n/2), 
\end{equation}
so that  $\Gamma_{\underline{m}}$ are $T$ independent.

\section{Decay rate: results}

\subsection{One-dimensional system}

Before entering the analysis of the 2D case, we first comment the 1D system (uncoupled chains), and 
specifically the stability of the DW perturbation in the presence of the Hubbard-type perturbation $H_U$. 
In contrast to the prototype
interacting spinless models (see e.g. Ref.~\onlinecite{gopal15}), much less is known on the existence 
of the nonergodicity within the Hubbard model \cite{schreiber15,mondaini15,barlev16}, whereby the symmetry 
arguments may imply also the restriction on the MBL physics \cite{potter16}. Our formulation of
the DW decay, Eq.~(\ref{gamma}), allows some additional insight into the problem by considering the condition for
$\Gamma>0$ in a macroscopic disordered system. While the density of MB states entering Eq.~(\ref{gamma})
is continuous and dense (for $L \to \infty$),  matrix elements $\langle \underline{n} |F|\underline{m} \rangle$ do not
connect states with $E_{\underline {n}} \sim E_{\underline {m}}$, since the interaction is local, while 
degenerate states can appear asymptotically only  at large space separation. The interplay 
and proper treatment of related resonances is in the core
of the theory of single-particle localization \cite{anderson58,mott68,kramer93} and of the MBL 
question \cite{fleishman80,basko06,oganesyan07,vosk15,imbrie16}. 
 Let us consider in  Eq.~(\ref{f}) only  the dominant (most local) term,
\begin{equation}
F \sim \frac{2U}{\sqrt{L}} \sum_{mm's}  \tilde \chi_{mm'} \zeta_{mm'} \bigl[\varphi^\dagger_{m's} \varphi_{ms} 
- \varphi^\dagger_{ms} \varphi_{m's} \bigr] \tilde n_{m,-s}. \label{fred}
 \end{equation}
where $\tilde \chi_{mm'} = \chi_{mm'}^{mm} \sim \phi_{mm'}$. Following a simple argument  by Mott 
\cite{mott68} for 1D non-interacting disordered system,
single-particle energies on n.n. sites cannot be close, i.e.  $|\epsilon_{m+1}-\epsilon_m | > 2 t$. 
In the same way one can get for more distant neighbors \cite{mott68},
\begin{equation}
|\epsilon_{m +r}-\epsilon_m | >  2t ~\mathrm{exp}(-\xi (r-1)),
\end{equation}
where $\xi \sim \ln(W_0/W)$ is the effective inverse localization length (averaged over band 
for large enough disorder $W > W_0 \sim 2 t$ ).
On the other hand, $\tilde \chi_{m,m+r}$ also decays as  $\propto ~\mathrm{exp}(-\xi r ) $.
So at least for $U \ll t$ we get the answer qualitatively consistent with the nonergodicity of DW correlations, 
$\Gamma_{\underline {m}} =0$. 
On the other hand, large $U>U_c$ (going beyond simple perturbation approach) are expected to lead to an 
ergodic behavior of DW perturbation with  $\Gamma_{\underline {m}} >0$, although the actual transition is 
not yet explored in detail 
within the 1D disordered Hubbard model \cite{schreiber15,mondaini15}. 

\subsection{Coupled identical Hubbard chains}

The introduction of the inter-chain hopping $t' \neq 0$ in Eq.~(\ref{hub})
qualitatively changes the physics in the case of identical disorder in all chains. 
Without interaction, i.e. at $U=0$, the eigenstates are a product of localized function and perpendicular 
plane waves. For simplicity we consider a 2D system, so that
\begin{eqnarray}
H_0 |\phi_{mqs}\rangle &=&(\epsilon_m + \tilde \epsilon_q ) |\phi_{mqs}\rangle, \nonumber \\
|\phi_{mqs}\rangle &=& \frac{1}{\sqrt{N}} \sum_{lj} \phi_{ml} \mathrm{e}^{iqj}  c^\dagger_{ ljs} |0 \rangle =
\varphi^\dagger_{lqs} |0 \rangle,
\end{eqnarray}
where $\tilde \epsilon_q =- 2t' \cos q$ and $N$ is the number of chains.  The interaction mixes such states,
\begin{eqnarray}
H_U  = \frac{U}{N}\sum_{\substack{ mm'nn' \\  qkp}} \chi_{mn}^{m'n'} \varphi^\dagger_{n', k+q \downarrow } \varphi_{nk \downarrow } 
\varphi^\dagger_{m',p-q\uparrow} \varphi_{m p\uparrow}. \label{hub2d} 
\end{eqnarray}  
The essential difference to possible decay in 1D, Eq.~(\ref{gamma}), is that the 
inter-chain dispersion leads to a continuous spectrum of overlapping initial and final states, so that the 
matrix elements in FGR, Eq.~(\ref{gamma}), can have finite values. Assuming for the moment that we are dealing 
with a weak 
perturbation $U<t'$, the evaluation of Eq.~({\ref{gamma}) leads to an effective (Boltzmann) density of 
decay channels, i.e. the density of states ${\cal D}(\omega)$, where (at $\beta \to 0$) 
\begin{eqnarray}
{\cal D}(\omega) &=&\mu \tilde {\cal D}(\omega), \qquad \mu = (1-\bar n/2 )^2 \bar n^2/4, \nonumber \\
\tilde {\cal D}(\omega)&=& \frac{1}{N^3} \sum_{ kpq } \delta ( \omega - \tilde \epsilon_{p-q}
- \tilde \epsilon_{k+q} + \tilde \epsilon_{p} + \tilde \epsilon_{k}) , \label{dos}
\end{eqnarray}
with $\int d\omega \tilde {\cal D}(\omega)=1$.
Distribution ${\cal D}(\omega)$ depends linearly on $t'$  and has a form as shown in Fig.~1,
with a singularity at $\omega\sim 0$. It is nonzero within the interval $-8 t' < \omega  < 8t'$ with a width 
$\sqrt {\bar \omega^2} \sim \sqrt{8} t'$. 

\begin{figure}[htb]
\centerline{\includegraphics[width=0.5\textwidth]{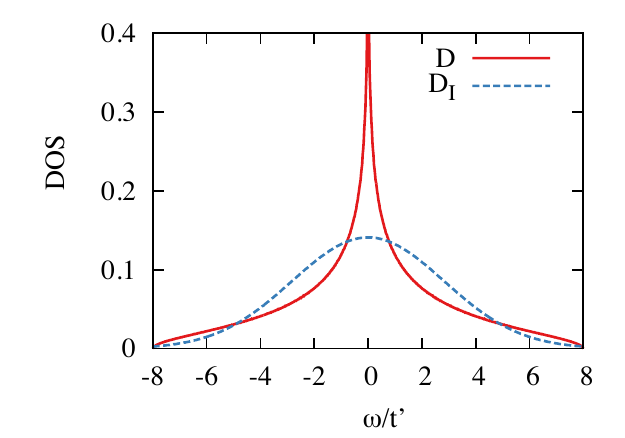}}
\caption{(Color online) Effective density of states $\tilde {\cal D}(\omega)$, emerging from the inter-chain hopping and entering
the evaluation of the decay rate $\Gamma$. Plotted is also the  incoherent approximant $\tilde {\cal D}_I(\omega)$.}
\end{figure}
Taking as the main contribution the reduced $F$, Eq.~(\ref{fred}), $\Gamma$ (at $\beta \to 0$) can be represented as
the sum of contributions emerging from different distances $r$, $\Gamma = \Gamma_1 + \Gamma_3 + \cdots$, where
\begin{equation}
\Gamma_r = 32 \pi \tilde \mu  U^2 |\tilde \chi_{m,m+r}|^2 \tilde 
{\cal D}(\Delta \epsilon_{r}=\epsilon_m - \epsilon_{m+r}), \label{gammar}
\end{equation}
and  $\tilde \mu = \mu/\langle B^2 \rangle=  \bar n (1-\bar n/2)/4$.

At least n.n. neighbors $r=1$ can be calculated more explicitly, taking into account the 
actual random distribution of $h_l$. Assuming for simplicity that we are dealing with a
two-level non-interacting problem with local potentials $h_l, h_{l+1}$, respectively, we get 
\begin{equation}
\Gamma_1 =32 \pi \tilde \mu  U^2 \int d \tilde h |\tilde \chi_{m,m+1}(\tilde h)|^2 
{\cal P}(\tilde h) {\cal D}( \Delta \epsilon_{1} (\tilde  h)), \label{gamma1}
\end{equation}
where $\tilde h = h_l - h_{l+1}$. In an analogous way one can treat also further neighbors $r\geq 3$, 
but here with an additional approximation that the effective in-chain hopping is reduced as $t_r 
\sim t (2t/W)^{r-1}$.
\begin{figure}
\centerline{\includegraphics[width=0.5\textwidth]{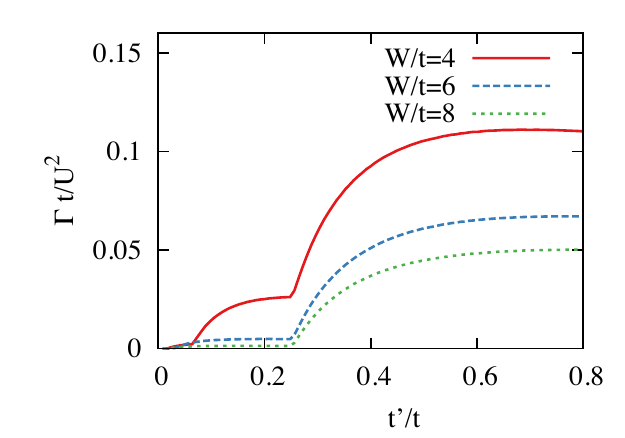}}
\caption{(Color online) Decay rate $ \Gamma t/ U^2$ vs. inter-chain  hopping $t'/t$ 
for different disorders  $W/t$ at fixed particle density $\bar n=1/2$.}
\end{figure}

The displayed result $\Gamma$ vs. $t'/t$, as shown in Fig.~2 is calculated  using Eq.~(\ref{gamma1}), 
at fixed $\bar n=1/2$ and  for various $W/t$. In spite of simplified approximations $\chi_{mn}^{m'n'}$ as well as 
for local energies $\epsilon_m$, several conclusions are straightforward: 

\noindent a) The decay rate becomes  $\Gamma>0$ for any finite 
$t' \neq 0$ and is proportional to $\bar n$ consistent with the origin in the interaction $U >0$ 
between fermions. $\Gamma \propto U^2$, at least within the perturbation regime considered 
analytically.

\noindent b) $\Gamma$ shows a steady increase with $|t'/t|$ for $t'/t<0.6$, consistent with 
experiments \cite{bordia16}. The decay rate $\Gamma$ vs. $t'/t$ is, at least within the 
approach used,  is not a simple function. Namely, for small $t'/t <0.2$~  $\Gamma$ is strongly reduced
since the contributions beyond the n.n. term $\Gamma_1$ become suppressed.
There appears also a saturation of $\Gamma$ for $t'/t > 0.6$. 
To some extent such behavior is plausible since excessively wide bands $t'/t>1$ 
cannot increase  $\Gamma$ much further.  

\subsection{Generalizations}

So far the analysis has been restricted to the regime of weak interaction $U/4 \ll  t,t'$,
whereby the factor four seems to be a fair estimate for the  crossover to a 
nonperturbative case. Since in the experiment \cite{bordia16} $t'/t$ is also varied, and 
of particular interest are results with $t'/t \ll 1$, one would wish to have an analytical result for the
intermediate regime  $ t' < U/4 < t$. If we consider in this case just the inter-chain part of the 
Hamiltonian, Eq.~(\ref{hub}) would for $U \gg t'$ transform into
\begin{equation}
H_\perp = \sum_l H_{l\perp},\qquad H_{l\perp} \sim - t' \sum_{js } ( \tilde c^\dagger_{l,j+1,s} \tilde c_{ljs} + \mathrm{H.c.}), \label{hproj}
\end{equation}
where $\tilde c_{ljs}  = c_{ljs} (1-n_{lj,-s} ) $ are projected fermion
operators.  Here, we omit possible exchange terms, since we are interested in systems with $\bar n <1/2$, i.e., away from half filling. 
As before the modified $H_\perp$ commutes with the DW operator, i.e. $[H_\perp,B] =0 $, hence it is
expected not to influence significantly the form of $F$, Eq.~(\ref{f}).  It is well
known \cite{ogata90} that eigenstates of the  projected model, Eq.~(\ref{hproj}), can be mapped on those 
of an noninteracting spinless model with the same single-particle dispersion $\epsilon_q=-2t' \cos q$. 
On the other hand, wavefunctions within the original basis are complicated and selection rules
changed. We therefore argue that within the intermediate regime the essential difference appears
in the evaluation of Eq.~(\ref{hub2d}), whereby the changed coherence factors between $q$-states and 
eigenstates of Eq.~(\ref{hproj}) lead to a different, rather incoherent ${\cal D}_I(\omega)$. For simplicity we
assume for the latter the Gaussian form with the same width $\bar \omega= \sqrt{8} t'$, i.e.,
\begin{equation}
{\cal D}_I(\omega)= \exp(-\omega^2/(4 t')^2)/\sqrt{16 \pi t'^2}.
\end{equation}
Taking ${\cal D}_I(\omega)$ 
as an input into Eqs.~(\ref{gammar}),(\ref{gamma1}) results are presented in Fig.~3. Results
differ from those in Fig.~2 only in some details. In particular, due to continuous ${\cal D}_I(\omega)$ 
the variation of $\Gamma$ vs. $t'/t$ is more gradual, but still showing a distinctive contributions
$\Gamma_{r>1}$ with strong $W$ dependence.

\begin{figure}
\centerline{\includegraphics[width=0.5\textwidth]{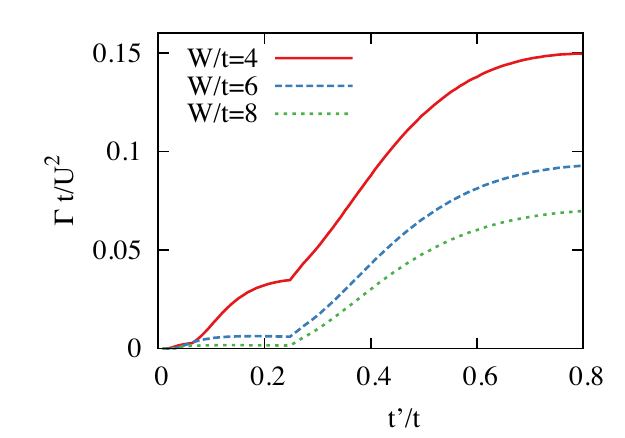}}
\caption{(Color online) Decay rate $\Gamma t/ U^2$ vs. $t'/t$ 
as calculated within the incoherent approximation 
for different $W/t$ at fixed $\bar n=1/2$.}
\label{fig1}
\end{figure}

The question of strong interactions $U>4t$ is more subtle.  One might employ an approximation
similar to Eq.~(\ref{hproj}) also for the in-chain terms, i.e.,
\begin{equation}
H_{j} \sim  - t \sum_{ls } ( \tilde c^\dagger_{l+1,js} \tilde c_{ljs} + \mathrm{H.c.}) 
+ \sum_l h_i n_{lj}. \label{hproj1}
\end{equation}
The message of such term is that the decay rate $\Gamma$ would not increase with $U>4t$, but would
saturate being finally determined by $t$, as emerging from Eq.~(\ref{hproj1}), as well as on
$t'$ and $W$. Taking strictly 1D model, as described by Eq.~(\ref{hproj1}), DW perturbation should not decay
at all due to the mapping on the spinless fermions and on the non-interacting Anderson model.  
Still, $t' \neq 0$ and the emerging 2D problem  does not have such a mapping,
so that inter-chain and in-chain  fermion states become coupled again.

\section{Conclusions}.
 
We presented a theory of a DW decay in the case of coupled disordered Hubbard
chains, with the identical disorder in each chain. It should be pointed out that we do not address the question 
whether the uncoupled 1D chains already show weak DW decay, but rather discuss the nontrivial additional 
contribution due to  the inter-chain coupling.  From the perturbation theory approach the decay
emerges due to Hubbard interaction $U>0$ mixing the  in-chain localized states and inter-chain
extended single-fermion states. The essential ingredient for $\Gamma>0$ (given by transition
rates between discrete localized states) are continuous 
spectra of overlapping extended states, i.e. with finite matrix elements . 
The latter are the the precondition for  an evaluation of $\Gamma$ within a FGR-type 
approximation. Taking into account that levels localized close in space are 
(on average) distant in  energy, this leads to quite strong dependence of $\Gamma$ on the ratio 
$t'/t$ as well as on an increase of $\Gamma$ with decreasing disorder $W$. The nontrivial structure within 
the dependence  on $t'/t$ emerges from a different regimes which allow for contributions beyond first n.n. in 
Eq.~(\ref{gammar}). The saturation of $\Gamma$ at $t' /t \sim 1$ is to some extent plausible since for $t' > t$ the
decay is limited by $t$ and not by $t'$, but can be also beyond the feasibility of initial assumptions. An 
interesting question is also to what extent the DW decay $\Gamma$ and possible MBL are sensitive to the difference of 
potentials in each  chain \cite{kasner88}, since even a small difference $\delta \epsilon > t'$ can induce also perpendicular
localization and prevent the DW decay  discussed above.

The theory is motivated by a concrete experiment on cold atoms \cite{bordia16}. We find that 
the variation of $\Gamma$, as measured via the time-dependent imbalance $I(\tau)$ with $U$ as well as on $t'/t$ and $W$ are qualitatively reasonably
reproduced. Still, several restrictions on the theoretical description should be taken into account. In
actual experiment a quasi-periodic (Aubry-Andr\`e) lattice is employed which is different from an
Anderson model with respect to the character and stability of localized states. Also, most results 
are available within the strong-interaction regime $U \gg 4 t$ where we cannot give an explanation 
on the  same level of validity, although the saturation (or a maximum) of $\Gamma$ for  $U > 4 t$ 
is expected.

\begin{acknowledgments}
The authors acknowledges the explanation of cold-atom experiments by P. Bordia and H. L\"uschen  
within the group of I. Bloch, LMU M\"unchen, and  fruitful discussions with F. Heidrich-Meisner and F. Pollmann. 
The author acknowledges also the support  of the Alexander von  Humboldt Foundation, 
as well the hospitality of the A. Sommerfeld Center for the Theoretical Physics, LMU M\"unchen, 
and the Max-Planck Institute for Complex Systems, Dresden, where this work has been started and major steps 
have been accomplished.  
\end{acknowledgments}

\bibliographystyle{apsrev4-1}
\bibliography{ref_manucmbl}

\end{document}